\documentstyle[twoside,fleqn,espcrc2,epsf]{article}
\newcommand{\noi}{\noindent}
\newcommand{\eq}{\begin{equation}}
\newcommand{\en}{\end{equation}}
\newcommand{\eqa}{\begin{eqnarray}}
\newcommand{\ena}{\end{eqnarray}}

\newcommand{\AmS}{{\protect\the\textfont2
  A\kern-.1667em\lower.5ex\hbox{M}\kern-.125emS}}
\title{
Lorentz gauge fixing and the Gribov problem: \\
the fermion correlator in lattice compact QED with Wilson fermions
\thanks{Support by EU TMR-network FMRX-CT97-0122, 
by grant INTAS-96-370, RFRB grant 99-01-01230 and by
JINR Heisenberg-Landau program is acknowledged}
}
 
\par
\vspace{1cm}

\author{
I.L. Bogolubsky$^{\rm a}$, V.K.~Mitrjushkin%
\address{Joint Institute for Nuclear Research, 141980 Dubna, Russia},
M. M\"uller-Preussker$^{\rm b}$,
P. Peter$^{\rm b}$ and N. V. Zverev%
\address{Humboldt-Universit\"at zu Berlin, Institut f\"ur Physik, 
Invalidenstr. 110, D-10115 Berlin, Germany}
}

\begin{document}

\begin{abstract}
For the Lorentz gauge the influence of Gribov copies on the
fermion propagator is investigated in quenched
lattice compact QED. In the Coulomb phase zero-momentum modes of the gauge
fields are shown to be the main reason for a significant deviation 
from ordinary perturbation theory. 
\vspace{1pc}
\end{abstract}

\maketitle

Numerical lattice computations of gauge-dependent 
gauge or fermion field correlators can give us detailed 
information about nonperturbative properties of
quantum fields and allow a direct comparison with perturbation theory
in the continuum. However, most of the known iterative gauge fixing 
procedures are faced with the existence of Gribov copies. 
For compact lattice QED within the Coulomb phase the
standard Lorentz gauge fixing causes
the transverse photon correlator to deviate significantly
from the expected zero-mass behaviour \cite{NaPl,BoMiMPPa}.
Also the fermion correlator depends strongly on the achieved gauge copies
\cite{NaSi}. Numerical \cite{BoMiMPPa,PhcrHetr,BoMiMPPe,BoMiDD} and
analytical \cite{Mitr} studies have shown that the main
gauge field excitations responsible for the occurence of disturbing
gauge copies are {\it double Dirac sheets} (DDS) and {\it zero-momentum
modes} (ZMM).

In the present talk we are going to discuss the influence of the ZMM 
on the fermion correlator. We shall demonstrate that only a proper 
account of them will allow us to determine the renormalized fermion mass.

We consider 4d compact QED in the quenched approximation
on a finite lattice ($V=N_s^3 \times N_t$).
We employ the standard Wilson (plaquette) gauge and fermionic action,
respectively. The latter is given by
\eq \label{2}   
S_F=\sum_{x,y}\overline{\psi}_{x}{\bf M}_{xy}(\theta)\psi_{y},
~~~{\bf M}={\bf 1} - \kappa {\bf D},
\en
$$
{\bf D}_{xy}=\sum_{\mu=1}^{4}
    \Bigl\{U_{x,\mu}P_{\mu}^{-}\delta_{y,x+\hat{\mu}}
          + U^{*}_{x-\hat{\mu},\mu} P_{\mu}^{+}\delta_{y,x-\hat{\mu}}\Bigr\},
$$
where $P_{\mu}^{\pm} = \hat{1} \pm \gamma_\mu$.
The $U_{x,\mu} = {\rm e}^{ {\rm i} \theta_{x,\mu}}$ 
denotes the gauge link degrees of freedom.
Periodic boundary conditions (b.c.) are implied. The fermion fields
are anti-periodic in $x_4$.
The fermion correlator for a given gauge field $~\theta~$ reads
\eq \label{4a}
\Gamma(\tau;\theta)=\frac{1}{V}\sum_{\vec{x},x_4}\sum_{\vec{y}}
{\bf M}^{-1}_{\vec{x},x_4; \vec{y},x_4 + \tau}(\theta).
\en
We shall restrict ourselves to the vectorial part
\eq \label{4b}
\Gamma_V(\tau;\theta) =\frac{1}{4}{\rm Re\,Tr\,}
\left(\gamma_4\Gamma(\tau;\theta)\right),
\en
with the trace taken with respect to the spinor indices.

Lateron, we shall compare the average
$~\langle~\Gamma_V~\rangle_{\theta}~$ within the approximation
where only constant gauge fields are taken
into account. The correlator in a uniform background 
$\theta_{x,\mu}\equiv\phi_{\mu},~-\pi<\phi_{\mu}\le\pi,$ $\mu=1,\cdots,4~$
is given by
\eqa \label{6b}
\Gamma_V(\tau;\phi)&=&
      \frac{(1-\delta_{\tau, 0}-\delta_{\tau, N_t}) (1+{\cal M})^{-1}}
             {2 \left(1+{\cal E}^{2N_t}-2{\cal E}^{N_t}\cos(\phi_4 N_t)\right)}
           \nonumber \\
&\times& \left([{\cal E}^{\tau}+{\cal E}^{2N_t-\tau}]\cos(\phi_4\tau) \right.
                     \\
&-&\left. [{\cal E}^{N_t+\tau}+{\cal E}^{N_t-\tau}]\cos[\phi_4(N_t-\tau)] 
                                                                      \right)
           \nonumber
\ena
where
\eqa
{\cal E}&=&\left[1+\frac{\sqrt{{\cal M}^2+{\cal K}^2}}{2(1+{\cal M})}
                                                                ~~\times\right.
            \nonumber \\
&&\left.\left(\sqrt{{\cal M}^2+{\cal K}^2}+\sqrt{({\cal M}+2)^2+{\cal K}
^2}\right)\right]^{-1},
            \nonumber
\ena
$$
{\cal M}=m_0+\sum_{l=1}^{3}\left(1-\cos \phi_l\right), \quad
{\cal K}=\sqrt{\ \sum_{l=1}^{3}\sin^2 \phi_l}.
$$
For $~\phi_{\mu}=0, ~~\mu=1,\cdots,4~$
the free fermion correlator for finite lattice size \cite{CaBa}
is reproduced. Note, that the bare mass $~m_0~$ is related to
the hopping-parameter by $~\kappa = 1/(8+2 m_0)$.

In numerical simulations the Lorentz gauge is fixed by
iteratively maximizing the functional
\eq \label{10}
F[\theta]=\frac{1}{4V}\sum_{x,\mu} \cos\theta_{x,\mu}=\mbox{Max.}
\en
with respect to (periodic) gauge transformations.

The algorithm is called standard Lorentz gauge fixing, if it consists
only of local maximization and overrelaxation steps. It is well-known
that this procedure gets stuck into {\it local}
maxima of the functional (\ref{10}) (gauge copies). 
It is a common belief that the Gribov problem has to be solved
by searching for the {\it global maximum} providing the best gauge copy.
In \cite{BoMiMPPe} we have shown that in order to reach the global maximum
we have necessarily to remove both the DDS and the ZMM from the gauge fields.

DDS can be identified as follows. The plaquette angle $~\theta_{x,\mu\nu}~$
is decomposed into the gauge invariant (electro-) magnetic flux
$~\overline{\theta}_{x,\mu\nu} \in (-\pi,\pi]~$
and the discrete gauge-dependent contribution
$~2 \pi n_{x,\mu\nu}, ~~n_{x,\mu\nu}=0,\pm 1,\pm 2.$ The latter
represents a Dirac string passing through the given plaquette if
$~n_{x,\mu\nu}=\pm 1~$ ({\it Dirac plaquette}). A set of Dirac
plaquettes providing a world sheet of a Dirac string on the dual
lattice is called {\it Dirac sheet}.
DDS consist of two sheets with opposite flux orientation
extending over the whole lattice and closing themselves
by the periodic b.c.
They can easily be identified by counting the total number of
Dirac pla\-quettes $N^{(\mu\nu)}_{DP}$ for each choice $(\mu;\nu)$.
DDS can be removed by periodic gauge transformations.
For standard Lorentz gauge fixing DDS have been seen to occur 
independently of the lattice size and $\beta$ 
\cite{BoMiMPPa,BoMiMPPe,BoMiDD}.
As a consequence the non-zero momentum transverse photon correlator 
significantly deviates from the expected zero-mass behaviour 
\cite{BoMiMPPe,BoMiDD,Mitr}.

\noi  The ZMM of the gauge field
\eq \label{13}
\phi_{\mu}=\frac{1}{V}\sum_{x}\theta_{x,\mu}
\en
do not contribute to the pure gauge field action either.
For gauge configurations representing small fluctuations around
constant modes it is easy to see, that the global maximum of the functional 
(\ref{10}) requires $~\phi_{\mu}\equiv 0~$.
The latter condition can be achieved by non-periodic gauge transformations
$$
\theta_{x,\mu} \rightarrow \theta^{\ c}_{x,\mu}=c_{\mu}+\theta_{x,\mu}
\quad {\rm mod\ }2\pi, \quad c_{\mu} \in (-\pi,\ \pi].
$$

We realize a proper gauge fixing procedure as proposed
in \cite{BoMiMPPe}. The successive Lorentz gauge iteration steps are always
followed by non-periodic gauge transformations suppressing the ZMM.
Additionally we check, whether the gauge fields contain
yet DDS. The latter can be excluded by repeating the procedure with 
initial random gauges.
We call the combined procedure {\it zero-momentum Lorentz gauge} (ZML gauge).
It provides with very high accuracy the global maximum of the gauge
functional. The photon propagator perfectly agrees with the expected
perturbative result throughout the Coulomb phase \cite{BoMiMPPe}.

Our Monte Carlo simulations were carried out with a filter heat bath
method. In order to extract the pure ZMM effect, we first
apply the standard Lorentz gauge procedure modified by initial random gauges
in order to suppress DDS. We denote this modified standard
Lorentz gauge procedure by LG. We compare the latter with the ZML
gauge described above.
%
\begin{figure}[tbp]
\vspace{0.5cm}
\epsfxsize=7.5cm\epsfysize=7.2cm\epsffile{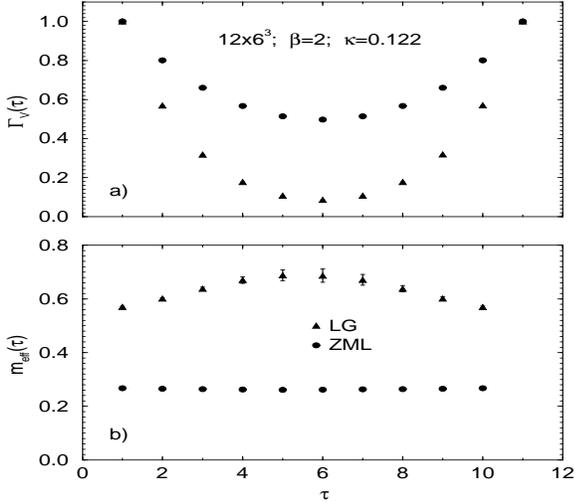}
\vspace{-0.5cm}
\caption{The fermion vector propagator ({\bf a}) and the effective
mass ({\bf b}) at $\beta=2$ and $\kappa=.122$ on a
$12\times 6^3$ lattice for LG and ZML gauges as explained in the text.}
\label{fig:fc_meff_12x06_b02p00_k122_2gau}
\end{figure}
For both these gauges we have computed the averaged fermion correlator 
employing the conjugate gradient method and point-like sources.
In the upper part of Fig. \ref{fig:fc_meff_12x06_b02p00_k122_2gau} we have
plotted $~\Gamma_V(\tau;\theta)~$
(normalized to unity at $\tau=1$). The situation seen is typical for
a wide range of parameter values within the Coulomb phase. Obviously,
there is a strong dependence of the fermion propagator  
on the gauge fixing procedure resulting in the presence or absence of ZMM.
The masses to be extracted seem to have different values. 
Let us determine the effective mass $~m_{eff}(\tau)$ in accordance with
\eq \label{16a}
\frac{\langle\Gamma(\tau+1;\theta)\rangle_{\theta}}
     {\langle\Gamma(\tau;  \theta)\rangle_{\theta}}
 =\frac{\cosh[E(\tau)(N_t/2-\tau-1)]}
       {\cosh[E(\tau)(N_t/2-\tau  )]}
\en
where $E(\tau)=\ln (m_{eff}(\tau)+1)$.
See the lower part of Fig. \ref{fig:fc_meff_12x06_b02p00_k122_2gau}.
In the LG case no plateau is visible, whereas
the ZML case provides a very stable one. 
Thus, only the ZML gauge yields 
a reliable mass estimate, whereas the standard method
to fix the Lorentz gauge obviously fails.
To get deeper insight into the effect of ZMM for the LG case 
(with DDS suppressed) we measure the probability distributions $~P(\phi)~$
for the space- and time-like components of ZMM according to Eq. (\ref{13}).
The distributions turn out to be flat up to an effective
cutoff at $~|\phi_{\mu}| \simeq \pi/N_{\mu}$ and to be
independent of $\beta$. In accordance with Eq. (\ref{6b})
we compute the fermion propagator for constant
modes in the LG case and average 
\eq
\langle\Gamma_V(\tau;\phi)\rangle_{\phi} =
\int [{\rm d} \phi]~ P(\phi)~ \Gamma_V(\tau;\phi).
\en

%
\begin{figure}[tbp]
\vspace{0.5cm}
\epsfxsize=7.7cm\epsfysize=7.2cm\epsffile{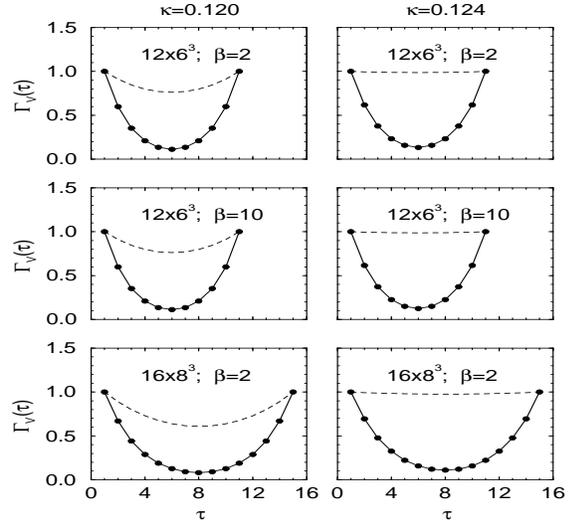}
\vspace{-0.5cm}
\caption{Free fermion propagator (dashed line)
and averaged constant-mode propagator in the LG case (full line)
for $\beta=2, 10$, $\kappa=.120, .124$, lattice size $12 \times 6^3$,
$16 \times 8^3$.
}
\label{fig:fc_tree_2lat_2bet}
\end{figure}
The results for several parameter sets are presented
in Fig. \ref{fig:fc_tree_2lat_2bet} together with the corresponding
free (i.e. zero-background) propagator.  

The model describes qualitatively the influence
of ZMM on the (full) fermion propagator in the LG case very well. 
It shows that the ZMM effect should not be expected to become weaker
with increasing $\beta$ and/or lattice size.
 

\end{document}